\documentclass[aps,tightenlines,showpacs]{revtex4}
\usepackage{graphicx}
\usepackage{float}
\newcommand{\bea}{\begin{eqnarray}}
\newcommand{\eea}{\end{eqnarray}}
\begin{document}
\title{\Large{Dielectric Functions and Dispersion Relations of 
Ultra-Relativistic Plasmas with Collisions}}
\author{ M.E. Carrington${}^{a,b}$, T. Fugleberg${}^{a,b}$, D. Pickering${}^{c}$ and M.H. Thoma${}^{d}$}
\email{carrington@brandonu.ca; fuglebergt@brandonu.ca; pickering@brandonu.ca; thoma@mpe.mpg.de}
\affiliation{ ${}^a$ Department of Physics, Brandon University, Brandon,
Manitoba,
R7A 6A9 Canada\\
${}^b$  Winnipeg Institute for Theoretical Physics, Winnipeg, Manitoba \\
${}^c$ Department of Mathematics, Brandon University, Brandon,
Manitoba,
R7A 6A9 Canada\\
${}^d$ Centre for Interdisciplinary Plasma Science,
Max-Planck-Institut f\"ur extraterrestrische Physik, P.O.Box 1312, 
D-85741 Garching, Germany }

\begin{abstract}
In the present paper we calculate the dielectric functions of an ultra-relativistic
plasma, such as an electron-positron or a quark-gluon plasma. We use classical 
transport theory and take into account collisions within the relaxation time 
approximation. From these dielectric functions we derive the dispersion
relations of longitudinal and transverse plasma waves.     
\end{abstract}


\maketitle

\section{Introduction}

In this paper, we will study the dielectric functions and dispersion relations for 
ultra-relativistic plasmas ($m \ll T$). The ultra-relativistic limit is relevant 
for the study of the quark-gluon plasma \cite{Mueller,QM} or a hot electron-positron 
plasma, which is present in various astrophysical situations \cite{Serbeto} but might also be 
produced in the laboratory \cite{Tsytovich}. Dispersion relations are of interest because 
they describe the propagation of collective plasma waves and provide information about 
physical quantities like screening lengths, oscillation  frequencies, damping and
particle production rates, transport coefficients, and the equation of state
\cite{Hwa}.

Dielectric functions and dispersion relations of ultra-relativistic plasmas
have been calculated using transport theory \cite{Silin} and thermal
field theory \cite{Klimov,Weldon}.
The new aspect of the present investigation is that we take into 
account collisions. For this purpose we
start from the Boltzmann equation and use the relaxation time approximation
for the collision term together with Maxwell's equations. We follow 
the non-relativistic derivation  presented in Refs.\cite{Alexandrov,Clemmow}. 
Note that in a plasma, electromagnetic fields may be produced by external sources. These 
fields change the distribution and motion of the charged particles in the plasma, 
creating induced charges and currents, which also produce electromagnetic fields. 
This system must be analysed self-consistently.

\section{Maxwell's Equations}

To obtain dispersion relations we start from Maxwell's equations in the plasma ($c=1$). For this purpose, 
we introduce the induced carge and current density $(\rho, j_i)$, the complex conductivity  $\sigma_{ij}$, 
and the dielectric tensor $\epsilon_{ij}$. Assuming a homogeneous medium these quantities are related to the 
electric field $E_i$ and electric displacement $D_i$ in momentum space by  
\bea
\label{defineEpsMom}
&& j_i(\omega,\vec k) = \sigma_{ij}(\omega,\vec k)E_j(\omega,\vec k),\\[2mm]
&& D_i(\omega,\vec k) = \epsilon_{ij}(\omega,\vec k)E_j(\omega,\vec k).\nonumber
\eea
In addition we note that the conductivity and the dielectric tensor are related by
\bea
\label{epsdefn}
\epsilon_{ij}(\omega,\vec k) = \delta_{ij}+\frac{i}{\omega}\sigma_{ij}(\omega,\vec k).
\eea
Using these definitions, in the absence of external sources, Maxwell's equations  
in momentum space become 
\bea
\label{maxwell2}
&& (\vec k \times \vec B)_i = -\omega\, \epsilon_{ij}(\omega,\vec k)\, E_j,\\[2mm]
&&\vec k \cdot \vec B =0,\nonumber \\[2mm]
&&(\vec k \times \vec E)_i = \omega \, B_i,\nonumber \\[2mm]
&& k_i \epsilon_{ij}(\omega,\vec k)E_j = 0. \nonumber
\eea
Eliminating $\vec B$ we obtain,
\bea
\left(k^2 \delta_{ij}-k_i k_j - \omega^2 \epsilon_{ij}\right) E_j = 0.
\eea
We further specialize to an isotropic medium where the dielectric tensor can be decomposed:
\bea
\label{defineEpsT}
&&\epsilon_{ij}(\omega,\vec k) = P^T_{ij}\epsilon^t(\omega,\vec k)+P^L_{ij}\epsilon^l(\omega,\vec k),\\[2mm]
&&P^T_{ij} = \delta_{ij}-\frac{k_ik_j}{k^2}\,,~~P^L_{ij} = \frac{k_ik_j}{k^2},\nonumber \\[2mm]
&& \epsilon^t(\omega,\vec k) = \frac{1}{2}P^T_{ij}\epsilon_{ij}\,,~~\epsilon^l(\omega,\vec k) = P^L_{ij}\epsilon_{ij}.
\eea
Using (\ref{defineEpsT}) we have,
\bea
\left(k^2-\omega^2\epsilon^t\right)\vec E_T=0\,,~~\left(\omega^2\epsilon^l\right)\vec E_L=0,
\eea
where
\bea
(E_T)_i = P^T_{ij} E_j\,,~~(E_L)_i = P^L_{ij} E_j.
\eea
Non-trivial solutions are obtained from 
\begin{eqnarray}
\label{disp1}
\epsilon^{l}(\omega ,k)&=&0, \\[2mm]
\omega^2 \epsilon^{t}(\omega ,k)-k^2&=&0.\nonumber
\end{eqnarray}

\section{Kinetic Theory and Dielectric Functions}

Let us consider an ultra-relativistic electron-positron plasma. We use kinetic theory 
to calculate the quantities $\epsilon^l$ and $\epsilon^t$ in (\ref{disp1}). The essence of kinetic 
theory is the assumption that the system can be described using a distribution function $f(\vec p,\vec r,t)$ 
which gives the probability density to find a particle of momentum $\vec p$ at position $\vec r$ at time $t$. 
In order to obtain an expression for $f(\vec p,\vec r,t)$ we must solve a kinetic equation. For the 
collisionless plasma it is common to use the  Vlasov equation:
\bea
\label{vlasov}
\frac{\partial f(\vec p,\vec r,t)}{\partial t}+ \vec v \frac{\partial f(\vec p,\vec r,t)}{\partial \vec r}
+ e\left[\vec E + \frac{1}{c}(\vec v \times \vec B)\right] \frac{\partial f(\vec p,\vec r,t)}{\partial \vec p}=0.
\eea
In order to solve this equation we make the approximation that the system is not far from equilibrium and 
expand around the equilibrium distribution function:
\bea
\label{deltaf}
f(\vec p,\vec r,t) = f_0(p) + \delta f(\vec p,\vec r,t)\,,~~~f_0(p) = \frac{4}{e^{p/T}+1}.
\eea
Note that we have neglected masses in the expression above by taking the ultra-relativistic limit ($m\ll T)$. 
The factor 4 in the equilibrium distribution $f_0(p)$ comes from the 2 spin states and from taking into account
electrons and positrons in (\ref{vlasov}) at the same time, which is possible since the final results depend
only on $e^2$. In addition we define the particle number density

\bea
N(\vec r,t) = \int dp f(\vec p,\vec r,t)\,,~~N_0 = \int dp f_0(p)\,,~~ dp:=\frac{d^3 p}{(2\pi)^3}.
\eea

In a weakly ionized plasma, we can include the effects of collisions by using a BGK collision term
\cite{Bhatnagar,Alexandrov,Clemmow}, i.e., we use a collision term on the right hand side of (\ref{vlasov}) of the form
\bea
\label{BGK}
-\nu\left[f(\vec p,\vec r,t)-N(\vec r,t) \frac{f_0(p)}{\int dp' f_0(p')}\right],
\eea
where $\nu$ is a velocity independent collision frequency. In perturbative QED the collision frequency between
electrons, positrons, and photons (elastic $2 \rightarrow 2$ scattering processes in the Boltzmann collision term) 
is of order $e^4$ (see e.g. 
\cite{Thoma}). Since we are interested in investigating the effect of the collision frequency on the dispersion 
relations (see below), we will vary the parameter $\nu$ from 0 and 1 in units of $m_\gamma$, which is the only scale other than $T$ 
in ultralrelativistic plasmas. (Note that $e\simeq 0.3$ is not far from 1). 

The BGK collision term corresponds to an improvement
of the relaxation time approximation for the collision term of the Boltzmann equation \footnote{In the BGK approximation
the current computed in transport theory is conserved, which is not the case with the relaxation time approximation.}. 
It simulates
the effect of close binary collisions with substantial momentum transfer \cite{Clemmow} and therefore  
it is particularly useful for describing collisions between ions and neutral gas molecules
in a weakly ionized plasma. However, it can also be used as a rough guide for describing
collisions between charged particles, such as in a hot electron-positron plasma. Note that
the effect of long distance collisions with small momentum transfer can be considered
by using a mean-field electric field $\vec E$ in (\ref{vlasov}) \cite{Pitaevskii}.

We substitute (\ref{deltaf}) into (\ref{vlasov}) and (\ref{BGK}) and linearize in the 
deviation from equilibrium $\delta f$ to obtain, 
\bea
\label{deltaf2}
\delta f(\vec p,\vec r,t) = \left[-i e \vec E \cdot \frac{\partial f_0(p)}{\partial \vec p}
+i\nu \eta f_0(p)\right] D^{-1}(p),
\eea
where
\bea
\label{eta}
\eta := \frac{1}{N_0}\int dp \,\delta f(\vec p,\vec r,t)\,,~~ D(p):= \omega +i\nu-\vec k \cdot 
\vec v(p)\,,~~\vec v(p) = \frac{\vec p}{p}.
\eea
Solving (\ref{deltaf2}) and (\ref{eta}) for $\delta f(\vec p,\vec r,t)$ we obtain an integral 
expression for the induced current:
\bea
j_i &&= e\int dp\, v_i(p)\, \delta f(\vec p,\vec r,t)\nonumber \\
&& =-ie^2\int dp  \,v_i(p)\,  \frac{(\vec E\cdot \vec p)}{p}\frac{\partial f_0}{\partial p}D^{-1}(p) \nonumber \\
&& + \frac{e^2\nu}{N_0}\int dp \,v_i(p) f_0(p) D^{-1}(p)
\int dp' \frac{(\vec E\cdot \vec p')}{p'} \frac{\partial f_0(p')}{\partial p'}D^{-1}(p')
\left[1-\frac{i\nu}{N_0}\int dp'' f_0(p'')D^{-1}(p'')\right]^{-1}. 
\eea
Using (\ref{defineEpsMom}) we extract an expression for $\sigma_{ij}(\omega,\vec k)$:
\bea
\label{theintegral}
\sigma_{ij} &&= -i e^2 \int dp \; v_i(p) v_j(p) \frac{\partial f_0(p)}{\partial p} D^{-1}(p) \nonumber \\
&& + \frac{e^2 \nu}{N_0}\int dp \,v_i(p) f_0(p) D^{-1}(p)
\int dp' v_j(p') \frac{\partial f_0(p')}{\partial p'}D^{-1}(p')
\left[1-\frac{i\nu}{N_0}\int dp'' f_0(p'')D^{-1}(p'')\right]^{-1}.
\eea
Using (\ref{epsdefn}) and (\ref{defineEpsT}) we obtain:
\bea
&&\epsilon^l = 1+\frac{i}{\omega k^2}(k_i k_j \sigma_{ij}),\nonumber \\
&&\epsilon^t = \frac{3}{2}+\frac{i}{2\omega}(\sigma_{ii})-\frac{\epsilon^l}{2}.
\eea

In the ultra-relativistic case we take $v=1$ which means that in  (\ref{theintegral}) the radial integral over $p$ and
the angular integrals decouple. Thus, in contrast to the non-relativistic case 
\cite{Alexandrov}, we are able to perform an analytic calculation of the dielectric functions. We define the integrals:
\bea
\label{integrals}
&& I_1 = \int_0^\infty dy\, \frac{y}{e^y+1} = \frac{\pi^2}{12},\nonumber \\
&& J_1 = \int ^1_{-1}dx \,\frac{x^2}{z-x} = - z\;X(z),\nonumber \\
&& J_2 = \int ^1_{-1}dx \,\frac{x}{z-x} =-X(z), \nonumber \\
&& J_3 = \int ^1_{-1}dx \,\frac{1}{z-x} = \frac{2-X(z)}{z},
\eea
where $z = (\omega+i\nu)/k$ and $X(z) = 2-z \ln\left(\frac{z+1}{z-1}\right)$.
Using the definitions of $f_0(p)$ and $D(P)$ given in (\ref{deltaf}) and 
(\ref{eta}) we obtain:
\bea
&& \epsilon^l = 1-\frac{2 e^2 T^2}{\pi^2 \omega k} I_1\left(J_1+\frac{i \nu}{2k}\,
\frac{J_2^2}{(1-\frac{i\nu}{2k}J_3)}\right), \nonumber \\
&& \epsilon^t = 1-\frac{e^2 T^2}{\pi^2 \omega k} I_1\left(J_3-J_1\right).
\eea
Substituting in from (\ref{integrals}) and defining the effective photon mass
$m_\gamma = e T/3$ we obtain:
\begin{eqnarray}
\label{disp2}
\epsilon^{l}(\omega ,k)&=&1+\frac{3m_\gamma^2}{k^2}\> 
\left (1-\frac{\omega +i\nu}{2k}\>
\ln \frac{\omega +i\nu +k}{\omega +i\nu -k}\right )\> \left (1-
\frac{i\nu}{2k}\> \ln \frac{\omega +i\nu +k}{\omega +i\nu -k}\right )^{-1},
 \\[2mm]
\epsilon^{t}(\omega ,k)&=&1-\frac{3m_\gamma^2}{2\omega (\omega +i\nu)}\> 
\left \{1+\left [\frac{(\omega +i\nu)^2}{k^2}-1\right ] \>
\left (1-\frac{\omega +i\nu}{2k}\> \ln \frac{\omega +i\nu +k}
{\omega +i\nu -k}\right )\right\}.
\nonumber 
\end{eqnarray}

Let us discuss the analytic structure of these dielectric functions by comparing them with the collisionless results.\\ 

\noindent (1). In the collisionless case, $\nu=0$, (\ref{disp2}) reduces to the expressions found by using the Vlasov equation \cite{Silin}, or the high-temperature
limit of thermal QED \cite{Klimov,Weldon}. The correction to the vacuum value $\epsilon^{l,t}=1$
is proportional to the square of the effective photon mass, i.e., of order $e^2T^2$.\\

\noindent (2). Note that the longitudinal dielectric function does not follow from the collisionless case by simply replacing
$\omega $ by $\omega +i \nu$. As a consequence, the longitudinal part of the polarization tensor, 
$\Pi^l=k^2(1-\epsilon^l)$ does not depend only on $(\omega +i\nu)/k$,  but also on $i\nu/k$ 
separately. The transverse part of the polarization tensor, on the other hand, $\Pi^t=\omega^2(1-\epsilon^t)$, depends only
on $(\omega +i\nu)/k$.\\

\noindent (3). Due to collisions, the dielectric functions (\ref{disp2}) have explicit imaginary parts from terms containing $i \nu$. In 
addition, even for $\nu=0$, there are imaginary parts coming from the logarithms, which are related to Landau damping.\\

Apart from the constant factor $m_\gamma$, (\ref{disp2}) agrees with the results for the dielectric
functions of a non-relativistic degenerate plasma as given in \cite{Alexandrov}. This agreement occurs because the radial and angular integrals 
decouple for both the relativistic and non-relativistic degenerate plasmas, and the angular integrals, giving the functional dependence on $\omega$
and $k$, are identical in both cases.
Note that for $\omega=0$, the longitudinal dielectric function becomes independent of the collision rate $\nu$.
Therefore, the (square of the) Debye screening mass $m_D^2=k^2[\epsilon^l(\omega=0)-1]=3m_\gamma^2$ is also 
independent of $\nu $.

\section{Dispersion Relations}

The equations (\ref{disp1}) and (\ref{disp2}) must be solved numerically. 
We scale all dimensionful variables by $m_\gamma$ and produce numerical results for 
$\omega/m_\gamma$ versus $k/m_\gamma$ for various values of $\nu/m_\gamma$. 
Figs. (1-4) show the real and imaginary parts of the dispersion relations for transverse 
and longitudinal modes. We correctly reproduce the known results from the Vlasov
equation (\ref{vlasov}) in the limit $\nu=0$ \cite{Silin}. These results can also be derived
from quantum field theory, using lowest order perturbation theory in the high-temperature
or hard thermal loop approximation \cite{Klimov,Weldon,Braaten}.           
The real part of the dispersion relation 
is remarkably insensitive to $\nu$. The imaginary part, which is identically zero when $\nu=0$, 
acquires a non-zero contribution for $\nu\ne 0$. 
%%%%%%%%%%%%%%%%%%%
\par\begin{figure}[H]
\begin{center}
\includegraphics[width=14cm]{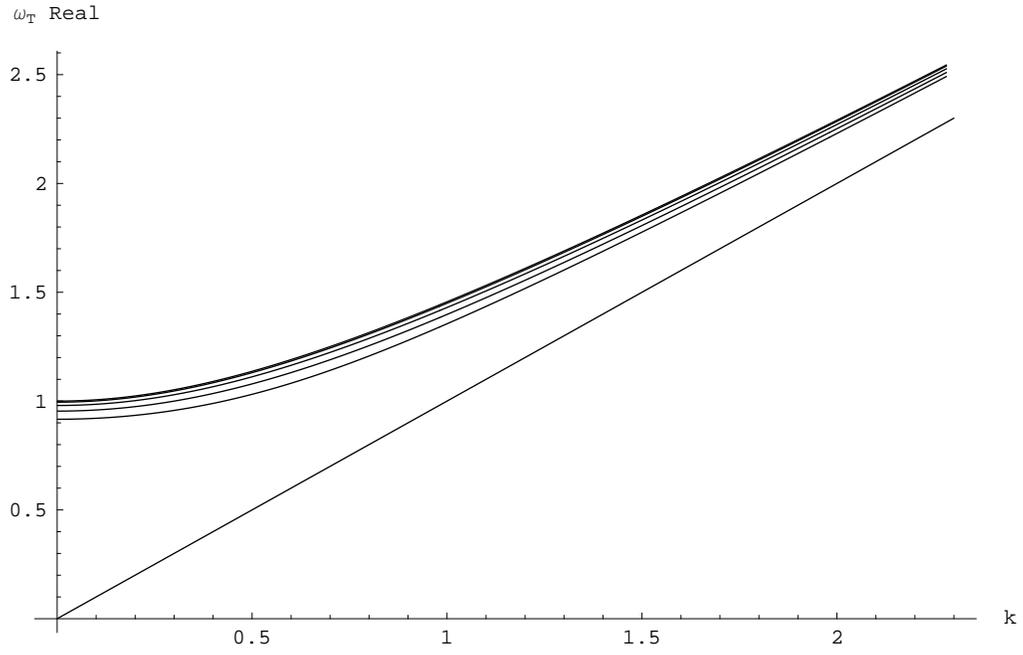}
\end{center}
\caption{Real part of the transverse dispersion relation (the parameter $\nu$ runs from 0 to 0.8 in steps of 0.2 with the
smallest value being at the top of the graph)}.
 \label{1}
\end{figure}
%%%%%%%%%%%%%%%%%%%%
%%%%%%%%%%%%%%%%%%%
\par\begin{figure}[H]
\begin{center}
\includegraphics[width=14cm]{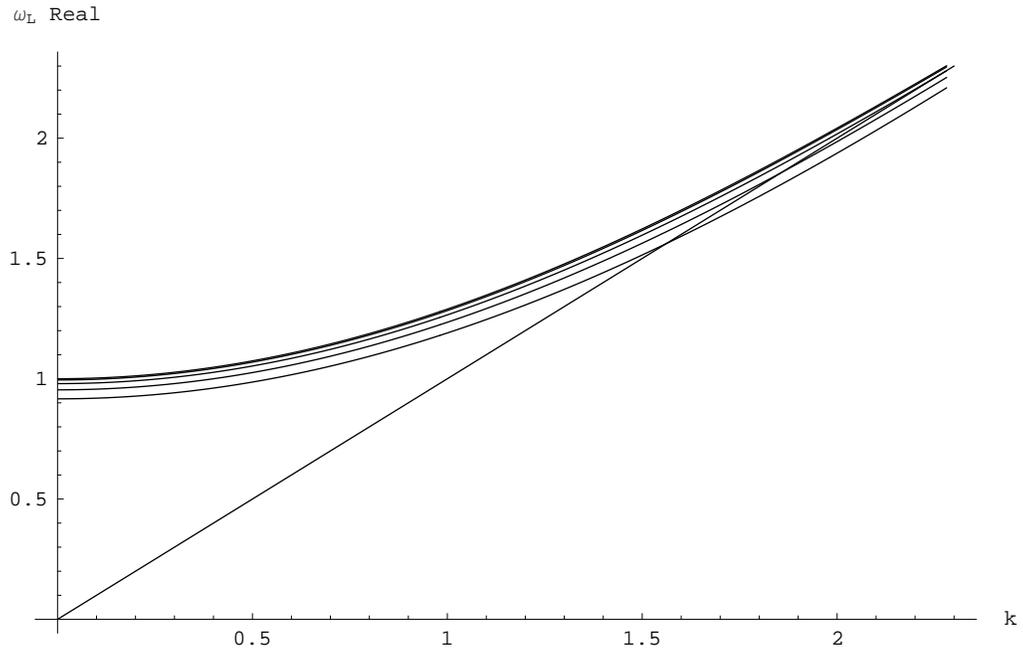}
\end{center}
\caption{Real part of the longitudinal dispersion relation (the parameter $\nu$ runs from 0 to 0.8 in steps of 0.2 with the
smallest value being at the top of the graph)}.
 \label{2}
\end{figure}
%%%%%%%%%%%%%%%%%%%%
%%%%%%%%%%%%%%%%%%%
\par\begin{figure}[H]
\begin{center}
\includegraphics[width=14cm]{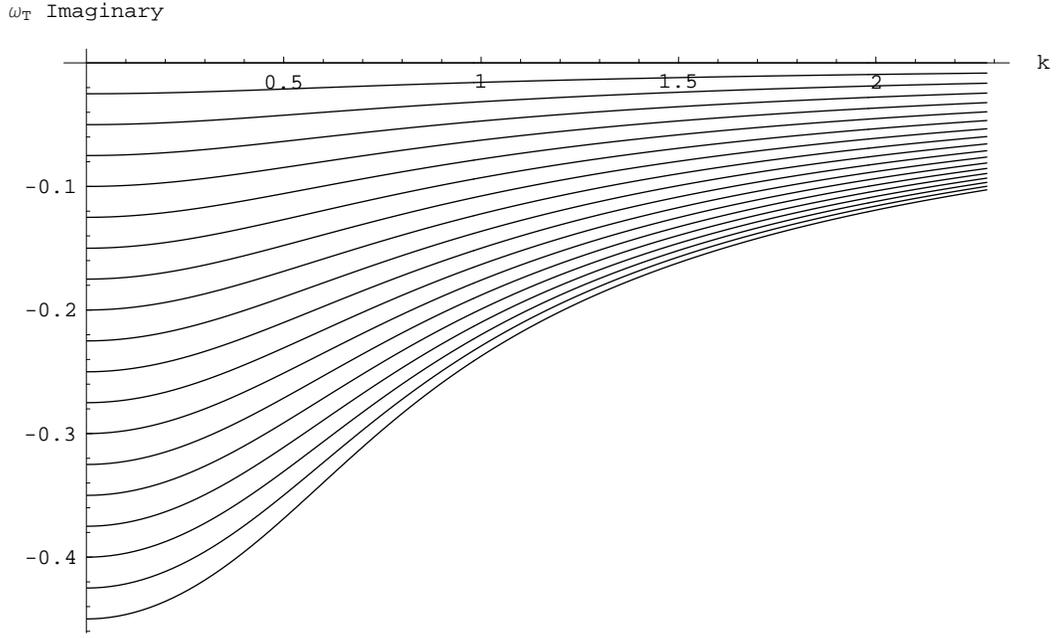}
\end{center}
\caption{Imaginary part of the transverse dispersion relation (the parameter $\nu$ runs from 0 to 0.9 in steps of 0.05 with the
smallest value being at the top of the graph)}.
 \label{3}
\end{figure}
%%%%%%%%%%%%%%%%%%%%
%%%%%%%%%%%%%%%%%%%
\par\begin{figure}[H]
\begin{center}
\includegraphics[width=14cm]{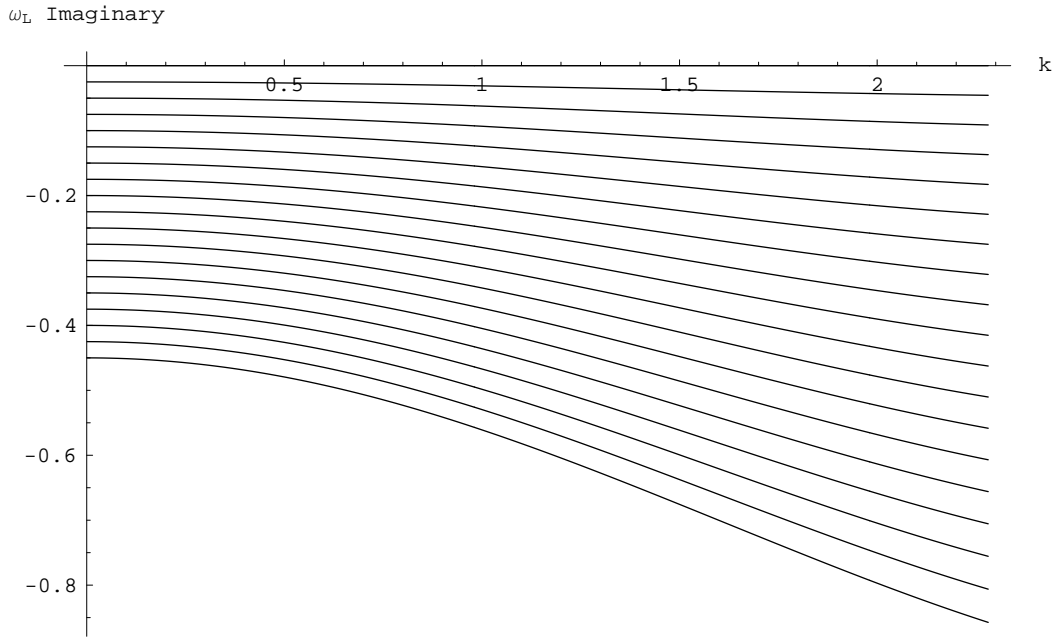}
\end{center}
\caption{Imaginary part of the longitudinal dispersion relation (the parameter $\nu$ runs from 0 to 0.9 in steps of 0.05 with the
smallest value being at the top of the graph)}.
 \label{4}
\end{figure}
%%%%%%%%%%%%%%%%%%%%
\section{Discussions and Conclusions}

In the present investigation we have calculated the dielectric functions and the dispersion relations of electromagnetic
plasma waves in an ultra-relativistic electron-positron plasma, taking collisions into account. Adding the BGK collision
term to the Vlasov equation, we have 
derived analytic expressions for the longitudinal and transverse dielectric functions in the case of a small
deviation from equilibrium. Using Maxwell's equations, the dispersion relations of longitudinal and transverse
plasma waves follow numerically from the dielectric functions. Surprisingly, the deviations from the collisionless
case are rather small, resulting only in a small shift of the dispersion relations towards lower energies. For example,
the plasma frequency, given by $\omega_L (k=0)=\omega_T(k=0),$ is reduced from $m_\gamma$ at $\nu =0$ to only 
about $0.9\> m_\gamma$ at $\nu =m_\gamma$ \cite{Schulz}. However, when collisions are included, the longitudinal dispersion at $\nu\neq 0$ intersects the light
cone $\omega =k$, in contrast to the case of collisionless dispersion, where $\omega_{L,T}(k)>k$ for all $k$. In the collisionless
case the only damping mechanism is Landau damping, which occurs for $\omega^2<k^2$, therefore plasma waves are undamped 
for $\nu =0$. For $\nu \neq 0$ the
plasma waves are damped by the collisions, which produces damping rates $\gamma_{L,T}(k) = -{\rm Im} \> \omega_{L,T}(k)$. The
longitudinal wave is also Landau damped for momenta that satisfy $\omega_L(k)<k$.   

It should be noted that we have introduced the collision rate $\nu$ phenomenologically as a constant free parameter. 
In hydrodynamical models one normally assumes that local equilibrium is achieved after some momentum independent
relaxation time \cite{Steffen}, which can be identified with the inverse of the collision rate.
In principle, the collision rate could be determined perturbatively leading to a momentum dependent rate proportional 
to $e^4$ \cite{Thoma}. However, it would be much more difficult to work with a momentum dependent collision rate. 

The effect of collisions in a quark-gluon plasma, e.g., on the damping of plasma oscillations and energetic quarks and gluons, 
has been discussed extensively 
in the literature using perturbative QCD as well as transport theory (see e.g. \cite{Thoma,Mrowczynski}).
Our results for the dielectric functions as well as the dispersion relations are valid for chromo-electromagnetic 
plasma waves in a quark-gluon plasma, if we replace the effective photon mass $m_\gamma$ by an effective gluon mass
$m_g=gT \sqrt{(1+n_f/6)/3}$, where $n_f$ is the number of quark flavors with masses $m\ll T$ \cite{Hwa}. However, this result 
is not fully consistent:  neglecting the non-abelian terms of order $g$ in the Yang-Mills equations contradicts the 
asumption that we can use a phenomenological collision term to model all collision processes, including higher order 
processes in $g$. At best, our phenomenological approach produces a rough 
estimate of the effect of collisions on plasma waves in a quark-gluon plasma.


\begin{thebibliography}{1}

\bibitem{Mueller} B. M\"uller, {\it The Physics of the Quark-Gluon Plasma}, Lecture Notes in Physics 225 (Springer, Berlin, 1985).
\bibitem{QM} For recent results on the quark-gluon plasma research see e.g. the recent ``Quark Matter'' proceedings in
Nucl. Phys. {\bf A715}, (2003) 3c, Nucl. Phys. {\bf A698}, (2002) 3c, and Nucl. Phys. {\bf A661}, (1999) 3c.
\bibitem{Serbeto} A. Serbeto, J.T. Mendoca, L.O. Silva, and P.K. Shukla, Phys. Lett. A {\bf 305}, 195 (2002). 
\bibitem{Tsytovich} V.N. Tsytovich and C.B. Wharton, Comments Plasma Phys. Controlled Fusion {\bf 4}, 91 (1978). 
\bibitem{Hwa} M.H. Thoma, in: {\it Quark-Gluon Plasma 2}, R.C. Hwa (Ed.) (World Scientific, Singapore, 1995); J.P. Blaizot and E. Iancu,
Phys. Rep. {\bf 359}, 355 (2002). 
\bibitem{Silin} V.P. Silin, Sov. Phys. JETP {\bf 11}, 1136 (1960).
\bibitem{Klimov} V.V. Klimov, Sov. Phys. JETP {\bf 55}, 199 (1982).
\bibitem{Weldon} H.A. Weldon, Phys. Rev. D {\bf 26}, 1394 (1982).
\bibitem{Bhatnagar} P.L. Bhatnagar, E.P. Gross, and M. Krook, Phys. Rev. {\bf 94}, 511 (1954).
\bibitem{Alexandrov} A.F. Alexandrov, L.S. Bogdankevich, and A.A. Rukhadze, {\it Principles of Plasma Electrodynamics} (Springer, Berlin, 1984).
\bibitem{Clemmow} P.C. Clemmow and J.P. Dougherty, {\it Electrodynamics of Particles and Plasmas} (Addison-Wesley, Reading, 1969). 
\bibitem{Thoma} M.H. Thoma, Phys. Rev. D {\bf 51}, 862 (1995).
\bibitem{Pitaevskii} L.P. Pitaevskii and E.M. Lifshitz, {\it Physical Kinetics} (Pergamon Press, Oxford, 1981).
\bibitem{Braaten} E. Braaten and R.D. Pisarski, Nucl. Phys. B {\bf 337}, 569 (1990).
\bibitem{Schulz} A reduction of the plasma frequency by higher order effects has also be found in thermal field theory by
H. Schulz, Nucl. Phys. B {\bf 413}, 353 (1994). 
\bibitem{Steffen} F.D. Steffen and M.H. Thoma, Phys. Lett. B {\bf 510}, 98 (2001).
\bibitem{Mrowczynski} S. Mr\'owczy\'nski, Phyys. Rev. D {\bf 39}, (1989) 1940.

\end{thebibliography}
\end{document}